\documentclass{PoS}
\usepackage{subfigure}

\PoS{PoS(LAT2005)118}

\newcommand{\dtau}{\Delta\hspace{-.06cm}\tau}

\title{HMC algorithm with multiple time scale integration and mass preconditioning }

\ShortTitle{HMC algorithm with multiple time scale integration and mass preconditioning }

\author{Karl Jansen\\
        John von Neumann Institute for computing, NIC, \\
        Platanenallee 6, D- 15738 Zeuthen, Germany\\
        E-mail: \email{Karl.Jansen@desy.de}}

\author{Andrea Shindler\\
        John von Neumann Institute for computing, NIC, \\
        Platanenallee 6, D- 15738 Zeuthen,  Germany\\
        E-mail: \email{Andrea.Shindler@desy.de}}

\author{\speaker{Carsten Urbach}\\
        Fachbereich Physik, Freie Universit{\"a}t Berlin, \\
        Arnimallee 14, 14195 Berlin, Germany\\
        E-mail: \email{Carsten.Urbach@physik.fu-berlin.de}}

\author{Urs Wenger\\
        John von Neumann Institute for computing, NIC, \\
        Platanenallee 6, D- 15738 Zeuthen,  Germany\\
        E-mail: \email{Urs.Wenger@desy.de}}

\abstract{We describe a new HMC algorithm variant we have
  recently introduced and extend the published results by 
  preliminary results of a simulation with a pseudo scalar mass value
  of $m_\mathrm{PS}\approx300\ \mathrm{MeV}$. This new run confirms
  our expectation that simulations with such pseudo scalar mass values
  become feasible and affordable with our HMC variant. In addition we
  discuss simulations from hot and cold starts at
  $m_\mathrm{PS}\approx300\ \mathrm{MeV}$, which we performed in order
  to test for possible meta-stabilities.}

\FullConference{XXIIIrd International Symposium on Lattice Field Theory\\
                 25-30 July 2005\\
                 Trinity College, Dublin, Ireland}

\begin{document}

\section{Introduction}

Even though Wilson's original discretization of the Dirac operator
gives rise 
to one of the clearest and best understood formulations of lattice
QCD, it shows problems in practice: due to explicit chiral symmetry
breaking the Wilson operator develops unphysically small eigenvalues,
which were thought to be responsible for instabilities observed in
dynamical simulations at light values of the quark masses with the
Hybrid Monte Carlo (HMC) algorithm \cite{Duane:1987de}.

However, recently it was discovered that -- rather surprisingly -- stable
simulations with the HMC algorithm are possible with values of the pseudo
scalar mass $m_\mathrm{PS}$ as low as $380\ \mathrm{MeV}$
\cite{Luscher:2004rx,Urbach:2005ji}, if a clever combination of
fermion determinant preconditioning and multiple time scale
integration is used\footnote{We expect that determinant
  preconditioning with the n-th root trick \cite{Clark:2004cq}
  performs similarly well.}. 
Moreover, the computational costs appear to 
be affordable, even with $m_\mathrm{PS}\approx 300\ \mathrm{MeV}$, if
the available results for the computational costs are 
extrapolated to this value of $m_\mathrm{PS}$. 

In this proceeding we report on progress with the
HMC variant we introduced in ref.~\cite{Urbach:2005ji}.

\section{Mass preconditioning}

For simplicity we consider here $N_f=2$ mass degenerate flavors of
Wilson fermions with Wilson-Dirac operator $D_\mathrm{W}$ and the
Wilson plaquette gauge action $S_\mathrm{g}$. The lattice action (for
one flavor) reads
\begin{equation}
  \label{eq:action}
  S = S_\mathrm{g} + \sum_x\bar\psi(x)(D_\mathrm{W} + m_0)\psi(x)\, ,
\end{equation}
where $m_0$ is the bare quark mass. For convenience we also introduce 
the hopping parameter $\kappa=(2 a m_0 + 8)^{-1}$ and the hermitian
Wilson-Dirac operator $Q=\gamma_5D_\mathrm{W}$. 

The numerical integration in the molecular dynamics part of the
HMC algorithm \cite{Duane:1987de} is usually performed
by means of the leap-frog algorithm, which is reversible and area
preserving, properties that are needed for the HMC algorithm to
be exact. We refer to ref.~\cite{Urbach:2005ji} for details on how the
leap frog algorithm is generalized to multiple time scales. In that
reference we also detail how to generalize the so-called
Sexton-Weingarten (SW) improved integration scheme \cite{Sexton:1992nu}. 

While the HMC variant presented in ref.~\cite{Luscher:2004rx} is based
on domain decomposition preconditioning, our variant relies on the so-called
Hasenbusch acceleration or mass preconditioning
\cite{Hasenbusch:2001ne}. It was realized in
ref.~\cite{Hasenbusch:2001ne} that using the identity 
\begin{equation}
  \label{eq:factorization}
  \det Q^2 = \det\left(Q^2+\mu^2\right)\ \det\left(\frac{Q^2}{Q^2 +\mu^2}\right)
\end{equation}
with a mass shift $\mu$ can speed up the HMC algorithm, if each of the
two determinants on the r.h.s. of eq.~(\ref{eq:factorization}) is
treated by a separate pseudo fermion field $\phi_i$ and a
corresponding pseudo fermion action $S_{\mathrm{PF}_i}$. The
acceleration comes 
about for the following reason: the condition number of $Q^2+\mu^2$
and $Q^2/(Q^2+\mu^2)$ is reduced when compared to the condition number
of $Q^2$. A reduced condition number is expected to lead to a reduced
molecular dynamics force and hence allows for larger step sizes in the
integration. At the same time the inversion of $Q^2+\mu^2$ is -- due to
the mass shift -- much cheaper than the inversion of $Q^2$, altogether
leading to a net speed up.

The original idea was then to choose the mass shift $\mu$ such that
the condition numbers of $Q^2+\mu^2$ and $Q^2/(Q^2+\mu^2)$ become
approximately equal. The speed-up was observed to be around a factor
of two \cite{Hasenbusch:2002ai}. 

\section{HMC with multiple time scale integration and mass preconditioning}

Motivated by the successful combination of multiple time scale
integration and determinant preconditioning via domain decomposition
in ref.~\cite{Luscher:2004rx}, 
we explored in ref.~\cite{Urbach:2005ji} the idea of combining mass
preconditioning with multiple time scale integration. With mass
preconditioning the Hamiltonian for the HMC algorithm reads
\begin{equation}
  \label{eq:generalH}
  H = \frac{1}{2} \sum_{x,\mu} P_{x,\mu}^2 + \sum_{i=0}^k S_{\mathrm{PF}_i}\, .
\end{equation}
The strategy is then to tune
$\mu$ in eq.~(\ref{eq:factorization}) such that the more expensive the
computation of a certain $\delta S_{\mathrm{PF}_i}$ is,  
the less it contributes to the total force. The different parts
of the action can then be integrated on different time scales
$\dtau_i$ chosen according to their force magnitude $F_i$, guided by
$\dtau_i F_i=\mathrm{const}$ for all $i$. 

In ref.~\cite{Urbach:2005ji} we demonstrated that this idea proves to
be useful in practice: we compared the performance of our HMC
algorithm variant
to the variant of ref.~\cite{Luscher:2004rx} and to a plain HMC as
used in ref.~\cite{Orth:2005kq}. The simulations were done on
$24^3\times32$ lattices with $\beta=5.6$ and pseudo scalar masses of
$m_\mathrm{PS}=665\ \mathrm{MeV}$, $485\ \mathrm{MeV}$ and $380\
\mathrm{MeV}$ (runs $A$, $B$ and $C$). Details for the algorithm
parameters as well as results for several quantities such as the
plaquette expectation value or the vector mass $m_\mathrm{V}$ can be
found in 
ref.~\cite{Urbach:2005ji}. In addition to these published
results we have one more simulation point, corresponding to
$m_\mathrm{PS}=294\ \mathrm{MeV}$ \cite{Luscher:talk} (run $D$). Our
simulations at this point are still ongoing and the history of this
run is not yet long enough to be fully conclusive. Nevertheless, we
present here first performance results for this point.

The first important observation from our investigations is that for all
four aforementioned simulation points the preconditioning masses and
time scales can be tuned such that simulations are
stable. Examples for Monte Carlo histories of the plaquette
expectation value or $\Delta H$ can be found in
ref.~\cite{Urbach:2005ji}. 

In order to compare the performance of our HMC variant to other
variants  we have chosen two different measures. The first is the
\emph{performance figure} 
$\nu = 10^{-3}(2n+3)\tau_\mathrm{int}(P)$ as introduced in
ref.~\cite{Luscher:2004rx}. $\tau_\mathrm{int}(P)$ is the integrated
autocorrelation time of the plaquette and $n$ is the number of
integration steps for the physical operator $Q^2$ necessary for one
trajectory. $\nu$ represents the number of inversions of the
operator $Q$ in thousands needed in order to obtain one independent
configuration. It is clearly algorithm and machine independent, but it
does not account for the preconditioning overhead, which is at least
for our HMC variant not completely negligible. 

\begin{table}[t!]
  \centering
  \begin{tabular*}{.9\textwidth}{@{\extracolsep{\fill}}lcccc}
    \hline\hline
    $\bigl.\Bigr.$& $\kappa$ & $\nu$ $[$this work$]$& $\nu$\cite{Luscher:2004rx,Luscher:talk} & $\nu$\cite{Orth:2005kq}\\
    \hline
    $A$ & $0.15750$ & $0.09(3)$ & $0.69(29)$ & $1.8(8)$\\

    $B$ & $0.15800$ & $0.11(3)$ & $0.50(17)$ & $5.1(5)$\\

    $C$ & $0.15825$ & $0.23(9)$ & $0.62(23)$ & -\\
    
    $D$ & $0.15835$ & {\color{red}$0.35$} & $0.74(18)$ & -\\
    \hline\hline
  \end{tabular*}
  \caption{Comparison of $\nu$ values from this work,
    ref.~\cite{Luscher:2004rx} (with updates from \cite{Luscher:talk})
    and ref.~\cite{Orth:2005kq}.} 
  \label{tab:nu}
\end{table}

The results for the $\nu$-values are summarized in table \ref{tab:nu}
and, while the $\nu$-values for our HMC variant and the one of
ref.~\cite{Luscher:2004rx,Luscher:talk} are compatible, they are
significantly smaller than the values extracted for the plain HMC
algorithm used in  ref.~\cite{Orth:2005kq}. Note that our $\nu$-value
for simulation point $D$ (in red) is only based on an extrapolation of
$\tau_\mathrm{int}(P)$ in $1/m_\mathrm{PS}^2$ and therefore preliminary.

\begin{figure}[t]
  \centering
  \subfigure[Comparison to ref.~\cite{Orth:2005kq}.]
  {\label{fig:berlin-walla} \includegraphics[width=.45\linewidth]
    {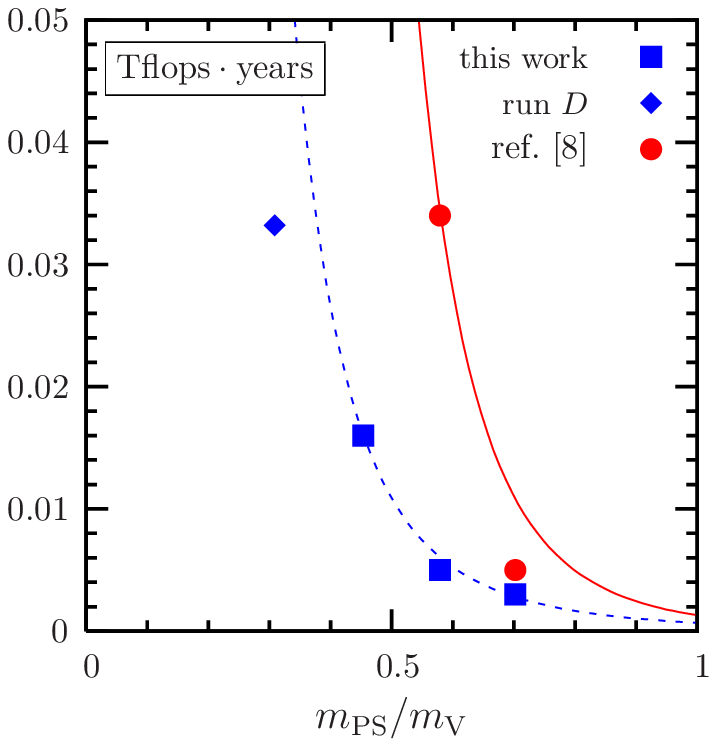}}
  \subfigure[Update of the ``Berlin Wall'' plots of
  refs.~\cite{Ukawa:2002pc,Jansen:2003nt}.]
  {\label{fig:berlin-wallb} \includegraphics[width=.443\linewidth]
    {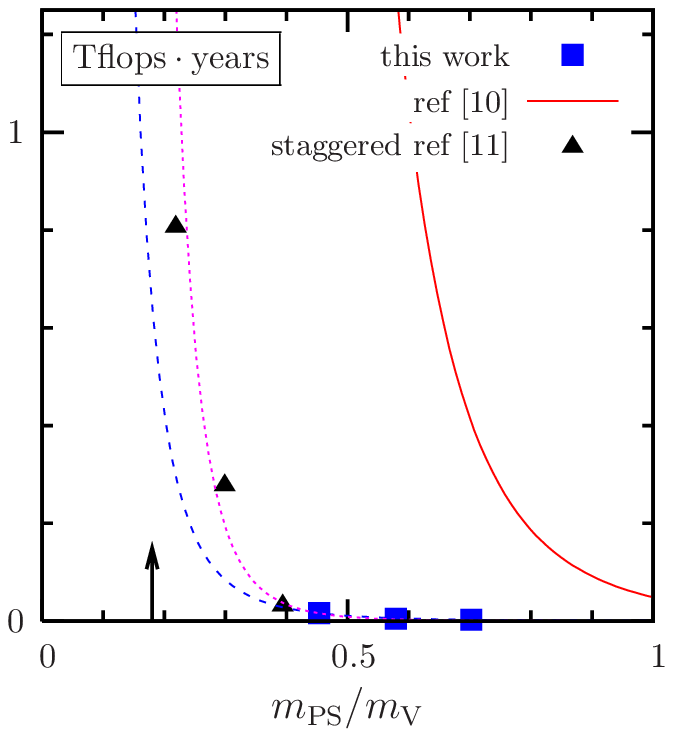}}
  \caption[Computer resources as a function of the quark mass ``Berlin
  Wall'' figure.]
  {Computer resources needed to generate $1000$ independent
    configurations of size $24^3\times 40$ at a lattice spacing of about
    $0.08\ \mathrm{fm}$ in units of $\mathrm{Tflops}\cdot
    \mathrm{years}$ as a function of $m_\mathrm{PS}/m_\mathrm{V}$. (See
    text for details.)}
  \label{fig:berlin-wall}
\end{figure}

The second performance measure we used is the number of floating point
operations (flops) needed to generate $1000$ independent
configurations of size $24^3\times 40$ with $a\approx0.08\
\mathrm{fm}$. For this measure we could compare our HMC variant to
the cost formula of ref.~\cite{Ukawa:2002pc}
\begin{equation}
  \label{eq:cost}
  C=K \left(\frac{m_\mathrm{PS}}{m_\mathrm{V}}\right)^{-4}L^5 a^{-7}\, .
\end{equation}
The actual value of $K$ can be found in \cite{Ukawa:2002pc}. The
result of the comparison is shown in figure \ref{fig:berlin-wall} as
an update of the ``Berlin Wall'' figures of
\cite{Ukawa:2002pc,Jansen:2003nt}. In figure \ref{fig:berlin-walla}
we compare our results represented by 
squares to the results of ref.~\cite{Orth:2005kq} represented by
circles. The lines are functions proportional to
$(m_\mathrm{PS}/m_\mathrm{V})^{-4}$ (dashed) and
$(m_\mathrm{PS}/m_\mathrm{V})^{-6}$ (solid) with a coefficient such
that they cross the data points corresponding to the lightest 
pseudo scalar mass. The diamond represents the preliminary result
of simulation point $D$, where the values for $\tau_\mathrm{int}(P)$
and $m_\mathrm{V}$ are extrapolated.
    
In figure \ref{fig:berlin-wallb} we compare to the formula of eq.~\ref{eq:cost}
\cite{Ukawa:2002pc} (solid line) by extrapolating our data with
$(m_\mathrm{PS}/m_\mathrm{V})^{-4}$ (dashed) and with
$(m_\mathrm{PS}/m_\mathrm{V})^{-6}$ (dotted), respectively. The arrow
indicates the physical pion to rho meson mass ratio. Additionally,
we add points from staggered fermion simulations as were used for the
corresponding plot in ref.~\cite{Jansen:2003nt}.
Note that all the cost data were scaled to match a lattice
time extend of $T=40$.

The most important conclusion from figure \ref{fig:berlin-wall} is
that with our HMC variant the ``Wall'' is substantially shifted
towards smaller values of the quark mass and that simulations with
Wilson fermions and $m_\mathrm{PS}\lesssim300\ \mathrm{MeV}$ become
feasible. Although the result for simulation point $D$ is preliminary,
it nicely confirms the results for larger values of $m_\mathrm{PS}$,
even under the pessimistic assumption that the final value might be a
factor of two larger.

\section{Thermalization property or meta-stability?}

Dynamical Wilson fermion simulations show the generic property of a
first order phase 
transition at the chiral point, as was shown in
ref.~\cite{Farchioni:2004us}. At this phase transition point the
PCAC quark mass jumps from non-vanishing negative to positive
values (or vice versa) and the pseudo scalar mass assumes a non-zero
minimal value, which can be rather large. This minimal value is
supposed to vanish as $a^2$ towards the continuum limit, but a
reliable information at which value of $a$ it takes a value below, say,
$300\ \mathrm{MeV}$, is missing. In ref.~\cite{Farchioni:2005tu} this
value of $a$ was estimated to be around $0.1-0.07\ \mathrm{fm}$. 

Since simulation point $D$ has $m_\mathrm{PS}\approx 300\
\mathrm{MeV}$ and the value of $a$ lies in the aforementioned interval,
it is important to investigate whether at this simulation point a
meta-stability is observed. To this end we performed for simulation
point $D$ two simulations, one started from an ordered and the other
from a disordered configuration. Both of these two runs reached now a
Monte Carlo history of about $1000$ trajectories, but it is still not
completely clear whether the runs have thermalized. 

Nevertheless, when measuring the PCAC quark mass for both runs during the
thermalization we observe that the run which started from a disordered
configuration shows a positive value of this quantity while the other
run has a negative value, indicating a meta-stability as observed in
ref.~\cite{Farchioni:2004us}. Only after around $900$ trajectories the
results of both runs approach each other and seem to converge to a positive
value of the quark mass. Hence, it seems that at these parameter values
no meta-stability occurs and the observed signs are simply thermalization
effects. Nevertheless, this observation emphasizes the importance of
checking for meta-stabilities \emph{before} large scale simulations
are started. It might also indicate that simulation point $D$ is close to a first
order phase transition that possibly occurs at lower values of
$m_\mathrm{PS}$.

\section{Conclusion}

In this proceeding we have reported on our progress with a new variant of
the HMC algorithm, which we introduced in
ref.~\cite{Urbach:2005ji}. The performance of our variant is
comparable to the recently introduced HMC variant with domain
decomposition \cite{Luscher:2004rx} and clearly superior to a plain
HMC algorithm. We presented an update of the ``Berlin Wall'' figure of
refs.~\cite{Ukawa:2002pc,Jansen:2003nt} showing that with our HMC
variant simulations with $m_\mathrm{PS}\approx 300\ \mathrm{MeV}$
become affordable and do not suffer from instabilities.

Moreover, we presented results of a check for meta-stabilities at our
simulation point with the lowest value of $m_\mathrm{PS}$. We observed
signs for a meta-stability during thermalization, which disappear only
after around $1000$ trajectories.

For the future it would be interesting to understand why the two HMC algorithm
variants -- the ones discussed here and in ref.~\cite{Luscher:2004rx}
-- allow for stable simulations with values of the pseudo scalar mass
of about $300\ \mathrm{MeV}$. One speculation is that this is due to the
infrared regularization of the operator spectrum provided by both,
mass and domain decomposition preconditioning. Another speculation is 
that the determinant and the forces are not well enough estimated with only
one pseudo fermion field, leading to possibly large fluctuations in the
forces. These fluctuations can be reduced by introducing additional
pseudo fermion fields. 

Clearly the clarification of these possibilities would be very
interesting and it might provide important insight to even further
improve the HMC algorithm.

\subsubsection*{Acknowledgments}

We thank I.~Montvay and I.~Wetzorke helpful discussions, the $qq+q$
collaboration and in particular F.~Farchioni, I.~Montvay and
E.E.~Scholz for providing us their analysis program for the masses and
C.~Destri and R.~Frezzotti for giving us access to a PC cluster
in Milano. We also thank the computer-centers at HLRN and at DESY
Zeuthen for granting the necessary computer-resources, and
M.~Hasenbusch for leaving us his HMC code as a starting point. This
work was supported by the DFG Sonderforschungsbereich/Transregio
SFB/TR9-03. 

\bibliographystyle{JHEP-2}
\bibliography{bibliography}

\end{document}